\documentclass[10pt,letterpaper]{article}
\usepackage{hyperref}
\usepackage{amsmath}
\usepackage{subfigure}
\usepackage{epsfig}

\begin{document}

%\twocolumn[
\title{\bf A class of line-transformed cloaks with easily-realizable constitutive parameters}

\author{Wei Xiang Jiang, Hui Feng Ma, Qiang Cheng and Tie Jun
Cui\footnote{E-mail: tjcui@seu.edu.cn}\\
\small \it State Key Laboratory of Millimeter Waves and Institute of Target Characteristics and Identification,\\
\small \it Department of Radio Engineering, Southeast University,
Nanjing 210096, P. R. China.}

\date{}
\maketitle

\begin{abstract}

We propose a class of line-transformed cylindrical cloaks which have
easily-realizable constitutive parameters. The scattering properties
of such cloaks have been investigated numerically for both
transverse-electric (TE) and transverse-magnetic (TM) incidences of
plane waves. A line-transformed invisibility cloak with a perfectly
electric conducting (PEC) inner boundary is actually a reshaping of
a PEC line to which the cloaked object is crushed. The numerical
results of near-field distributions and far-field scattering properties have
verified the above conclusions. We also investigate the relationship
between the constitutive parameters of a line-transformed cloak and
the length of the corresponding line. The changing range of constitutive
parameters is large when the line is short, while the changing range
becomes small when the line is long. The above conclusion
provides an efficient way to realize the invisibility cloaks using
artificial metamaterials.

\vskip 5mm


\vskip 5mm


\end{abstract}
%]

\newpage

\section{Introduction}

In the past three years, great attention has been paid to the
invisibility cloaks due to their fascinating and exciting properties
[1-23]. The interesting idea to control the electromagnetic (EM)
waves was proposed in Refs. [1] and [2]. The first free-space cloak
was experimentally demonstrated at the microwave frequency using the
reduced constitutive parameters [3]. Later, Li and Pendry have
proposed a kind of ground-plane cloak recently [4], which was
verified soon in the microwave regime [5] and in the optical regime
[6,7]. Many further analytical and numerical investigations have
also been devoted in the recently years [8-23].

To design a free-space cloak, one usually transforms an enclosed
space to an annular space with the outer boundary unchanged [1]. As
illustrated in Fig. 1(c), the region enclosed by boundary $s$ is
compressed to the annular region bounded by the outer boundary $s$
and an inner boundary $s_0$. Such a space mapping can be expressed
as a specific coordinate transformation. Generally, the inner
boundary $s_0$ is obtained by blowing up a line or a point as shown
in Figs. 1(a) and 1(b) [23]. Thus the invisibility cloaks can be
divided into two classes: point-transformed cloaks and
line-transformed cloaks. The green region in Fig. 1(c) is the
desired cloaking layer.

One problem for the point-transformed full-parameter cloaks is that
the parameters always exist singularities on the inner boundary. In
other words, some parameter components approaches infinity near the
inner boundary, which is difficult to reach even using modern
metamaterials [3]. Recently, the circular cloaks [14,16], elliptical
cloaks [17-20], and other-shaped cloaks [13,21,23] have been
designed and studied in different coordinate systems. However, the
material parameters for the above cloaks still have infinite values
on the inner boundary. Mathematically, such cloaks try to crush an
shielded region to a point, which leads to the singularity of
constitutive parameters.

In view of the difficulty to realize the point-transformed
full-parameter cloak, a ground-plane cloak which can hide the object
under a metamaterial carpet has been proposed [4]. Unlike the
completely free-space cloaks, the ground-plane cloak crushes the
hidden object to a conducting wire. Suck a carpet-like cloak does
not require extreme values of the material parameters. But the
carpet-like cloak can only shield object under a conducting surface,
and cannot shield objects in free space.

In this work, we propose a class of free-space line-transformed
cylindrical cloaks which have easily-realizable constitutive
parameters. Instead of shrinking the concealed object to a point as
usual, such cloaks crush the object to a line segment. The
scattering properties of such cloaks have been investigated
numerically for both transverse-electric (TE) and
transverse-magnetic (TM) incidences of plane waves. A
line-transformed invisibility cloak with a perfectly electric
conducting (PEC) inner boundary is indeed a reshaper of a PEC line
to which the cloaked object is crushed. Numerical results of
near-field distributions and scattering patterns have verified the
above conclusions. We also investigate the relationship between the
constitutive parameters of a line-transformed cloak and the length of
the corresponding line. The changing range of constitutive parameters is large
when the line is short, while the changing range becomes small when
the line is long. The proposed invisibility cloaks could be realized
using artificial metamaterials.

\section{Line-transformed cloaks and constitutive parameters}

Using the unique feature of classical elliptically-cylindrical
coordinate system, we construct a class of line-transformed cloaks.
The relationship between this coordinate system
$(\xi,~\eta,~z)$ and the Cartesian coordinates $(x,~y,~z)$ is
written as
\begin{eqnarray}
x=p\cosh \xi \cos \eta,\quad y=p \sinh \xi \sin \eta,\quad z=z,
\end{eqnarray}
in which $p$ is the half focus of the ellipse. We note that in this
coordinate system, if $p$ is assumed to be constant, then isolines
for $\xi$ can be a series of elliptical cylindrical shells with the
same focus value. In particular, $\xi=0$ means a line segment
$(y=0,~p\leq x\leq p)$ with length $2p$. To construct a
line-transformed cloak, a general spatial transformation from the
elliptical region $\xi\in[0, \xi_2]$ to the annular region
$\xi'\in[\xi_1, \xi_2]$ can be expressed as follows
\begin{eqnarray}
\xi'=f(\xi),~\eta'=\eta,~z'=z,
\end{eqnarray}
in which $f$ is a continuous and differentiable function with
$f(0)=a_1$ and $f(a_2)=a_2$, where $a_1$ and $a_2$ are coordinates of the
inner and outer boundaries of the cloak. The relationship between
coordinates and lengths of major axes can be expressed as
$a_1=\ln\big(r_1/p+\sqrt{(r_1/p)^2-1}\big)$ and
$a_2=\ln\big(r_2/p+\sqrt{(r_2/p)^2-1}\big)$, where $r_1$ and $r_2$
are the lengths of major axes for inner and outer shells of the
cloak. Obviously, the inner boundary is crushed to the line segment
$2p$ using the above coordinate transformation.

The constitutive tensors of the cloak in an arbitrary coordinate
system are given as the following equations
\begin{equation}
\overline{\varepsilon'}=\Lambda\overline{\varepsilon}\Lambda^{T}/\det(\Lambda),~~
\overline{\mu'}=\Lambda\overline{\mu}\Lambda^{T}/\det(\Lambda).
\end{equation}
Here, the Jacobian matrix is
\begin{equation*}
\Lambda=
\begin{pmatrix}
f'g & ~~~~0 &~~~~ 0\\
0&~~~~g&~~~~0\\
0&~~~~0&~~~~1 \end{pmatrix},
\end{equation*}
in which $f'$ is the derivative of $f$ with respect to $\xi$ and
$g=\sqrt{\cosh 2\xi'-\cos 2\eta}/\sqrt{\cosh 2\xi-\cos 2\eta}$. The
invisibility cloak is considered to be placed in the free space, in
other words,
$\overline{\varepsilon}=\varepsilon_0\overline{I},~\overline{\mu}=\mu_0\overline{I}$.
Then the relative permittivity and permeability tensors of the
line-transformed cloak are expressed as
\begin{eqnarray}
&&\varepsilon'_{\xi'}=\mu'_{\xi'}=f',\\
&&\varepsilon'_{\eta'}=\mu'_{\eta'}=\frac{1}{f'},\\
&&\varepsilon'_{z'}=\mu'_{z'}=\frac{1}{f'}\frac{\cosh 2f^{-1}-\cos
2\eta'}{\cosh 2\xi'-\cos 2\eta'},
\end{eqnarray}
where $f^{-1}$ denotes the inverse function of $f$. The cloak is
considered to be lossless in the working frequency. For the sake of
simplicity and real applications, we choose $f$ as a linear function
$f(\xi)=(a_2-a_1)\xi/a_2+a_1$ in the following discussions. As a result, we obtain
$f'=(a_2-a_1)/a_2$ and $f^{-1}(\xi')=a_2(\xi'-a_1)/(a_2-a_1)$.

Equations (4)-(6) provide full-design material parameters for a
general line-transformed cloak in the classical
elliptical-cylindrical coordinates. The relationship between the
permittivity and permeability components and the length of the line
segment $2p$ has been shown in Figure 2, in which we choose the
lengths of major axes as $r_1=0.2$ m and $r_2=0.4$ m. To visualize the
changing ranges of material parameters in cloaking layer better, we plot the
parameter components along the line $\eta'=0$. When the
half focus $p$ changes from 0 to $r_1$, the $\xi$ and $\eta$
components of the material parameters are shown in Fig. 2(a), and the $z$
component ranges from 0 to $\eta$ as shown in Fig. 2(b), which can
be easily observed from Eq. (6). Clearly, this set of material
parameters have no singularities because the line-transformed cloak
crushes the concealed object to a line segment with nonzero length
$2p$.

For practical reasons, we choose two cases $p=0.0476$ m and
$p=0.1464$ m to study the material properties of line-transformed
cloaks. The parameter components have been illustrated in Fig. 2(b).
When selecting $p=0.0476$ m, then we have $\varepsilon_{\xi}=\mu_{\xi}=0.25$,
$\varepsilon_{\eta}=\mu_{\eta}=4$, and the $z$ components range from 0
to 4. When $p=14.64$, then $\varepsilon_{\xi}=\mu_{\xi}=0.5$,
$\varepsilon_{\eta}=\mu_{\eta}=2$, and the $z$ components range from 0
to 2. In these two cases, the material properties in the cloaking
region are much easier to achieve using metamaterial structures. We
will investigate the cloaking performance of the line-transformed cloaks
for such two cases in the next section.

\section{Discussions and simulations}

To illustrate the invisible properties of the line-transformed
cloaks, we report the results of some accurately numerical
simulations. All simulations are performed by using the software
package, COMSOL Multiphysics, which is based on the finite element
method (FEM). Here, both TE-polarized and TM-polarized time-harmonic
incident plane waves will be considered. For the TE-wave incidence, only
$\mu_{\xi}$, $\mu_{\eta}$, and $\varepsilon_{z}$ components of the
material parameters are required for the simulations; for the TM-wave
incidence, only $\varepsilon_{\xi}$, $\varepsilon_{\eta}$, and
$\mu_{z}$ are of interest. The working frequency is chosen as 3 GHz, at which
the required material properties are easily realized by artificial
metamaterials. As mentioned above, in all simulations, the lengths
of major axes for inner and outer elliptical shells of the cloak are
set as 0.2 m and 0.4 m, respectively, and the two ellipses have the
common focus.

First, we consider the case when the line segment is
chosen as $2p=0.0952$ m and TE-polarized plane waves are
horizontally incident from the left to the right. Figure 3(b) shows the total
electric-field distribution of the line-transformed cloak, which
contains a PEC inner boundary. In Fig. 3(a), the electric-field
distribution of a PEC line is shown, in which the 'dashed line'
indicates the outer boundary of the cloaking layer used in the
design of the line-transformed cloak. We draw this boundary only for
the sake of visualization. By comparison of Figs. 3(a) and 3(b), we
observe that the scattering pattern of the line-transformed cloak with
PEC inner boundary is exactly the same as that of a PEC line. To
illustrate this better, we calculate the scattering widthes for the bare
PEC cylinder, the line-transformed cloak, and the corresponding PEC
line, as plotted in Fig. 3(c). The scattering widthes and
field distributions reveal that the line-transformed cloak with PEC
inner boundary is perceived as a PEC line when observed anywhere
outside the boundary of the cloaking layer.

When TM-polarized waves are incident along the horizontal direction,
the total magnetic-field distribution of the line-transformed cloak
with the PEC inner boundary is illustrated in Fig. 4(b). In such a
case, the magnetic-field distribution behave exactly the same as
that of a PEC line shown in Fig. 4(a). Evidently, a much smaller
scattering outside the cloaking region is observed in Figs. 4(a) and
4(b). The far-field scattering patterns for the cloak and PEC line are
exactly the same, as demonstrated in Fig. 4(c). Hence, a line-transformed
cloak with PEC inner boundary is reshaped as a PEC line scatterer.

Next we consider the case when the PEC line is chosen as $2p=0.2928$
m. Figure 5 illustrates the numerical results of electric fields in
the computational domain under the incidence of TE-polarized plane
waves. When the waves are incident along the horizontal direction,
the electric-field distributions of the line-transformed cloak with
the PEC inner boundary and the PEC line are shown in Figs. 5(a) and
5(b), from which we clearly see that the field values are exactly
the same. The comparison of scattering widthes is shown in Fig.
5(c). When the TM-polarized plane waves are incident, the
magnetic-field distributions behave significantly different from
those in Fig. 5, as demonstrated in Fig. 6. Clearly, the near-filed
and far-field distributions are exactly identical for the cloak and
the PEC line, as shown in Figs. 6(a), 6(b) and 6(c).

The above simulation results verify that a line-transformed cloak
with a PEC inner boundary scatters the waves exactly like a certain
PEC line for both TE-wave and TM-wave incidences. In other words, the
line-transformed cloak with the PEC inner boundary is indeed a
reshaper of the PEC line. Hence, for the line-transformed cloak with the
PEC inner boundary, the significant difference between TE-wave and
TM-wave incidences is similar to the case for a PEC line. From above
simulation results, we also observe that the line-transformed cloak
have strong scattering for the TE-wave incidence
(see Figs. 3(b) and 5(b)), but have weak scattering for the
TM-wave incidence (see Figs. 4(b) and 6(b)).
Such a effect can be explained from the viewpoint of the cloaking
shell which shrinks the PEC inner boundary to the PEC line. It is
well-known that a PEC line forces the tangential electric field to
zero, which can be observed in Figs. 3(a) and 5(a) for the TE incidence.
However, the PEC line can support a discontinuity of the tangential
magnetic field and the scattering decreases significantly for the TM
incidence, as shown in Figs. 4(a) and 6(a). Hence, for the TM
incident case, the cloaking performance is significantly better.

\section{Summary}

In summary, we have studied the scattering properties of a class of
line-transformed cloaks numerically for both TE and TM incidences of
plane waves. Such a cloak with the PEC inner boundary is indeed a
reshaping of a PEC line to which the shielded object is crushed. The
numerical results of near-field distributions and far-field
scattering patterns have confirmed the above conclusions. We have also
investigated the relationship between the material parameters of
the line-transformed cloak and the length of the corresponding line, which
provides an efficient way to realize the invisibility cloaks using
artificial metamaterials.

\section*{Acknowledgement}

This work was supported in part by the National Science Foundation
of China under Grant Nos. 60990320, 60990324, 60671015, 60871016,
and 60901011, in part by the Natural Science Foundation of Jiangsu
Province under Grant No. BK2008031, and in part by the 111 Project
under Grant No. 111-2-05. WXJ acknowledges the support from the
Graduate Innovation Program of Jiangsu Province under No.
CX08B\_074Z.

\newpage

\section*{{{\bf List of Figure Captions}}}

\noindent \textbf{Fig. 1:} {(color online) A scheme of the cloak's
inner boundary for (a) a point-transformed cloak and (b) a
line-transformed cloak. (c) The cross section of a general
cylindrical cloak (green region).}

\vskip 5mm

\noindent \textbf{Fig. 2:} {(color online) (a) Permittivity and permeability
components for the line-transformed cloak when the line segment $2p$
changes from 0 to $2r_1$. (b) Permittivity and permeability
components for two different cases when $p=0.0476$ m and $p=0.1464$ m.}

\vskip 5mm

\noindent \textbf{Fig. 3:} {(color online) TE-polarized plane waves
are incident along the horizontal direction when $p=0.0476$ m. The electric-field
distributions of (a) the line-transformed invisibility cloak with the PEC
inner boundary, (b) the PEC line. (c) The scattering width for the
line-transformed cloak, the PEC line and a bare PEC cylinder.}

\vskip 5mm

\noindent \textbf{Fig. 4:} {(color online) TM-polarized plane waves
are incident horizontally when $p=0.0476$ m. The magnetic-field distributions of (a) the
line-transformed invisibility cloak with the PEC inner boundary, (b) the
PEC line. (c) The scattering width for the line-transformed cloak, the
PEC wire, and the bare PEC cylinder.}

\vskip 5mm

\noindent \textbf{Fig. 5:} {(color online) TE-polarized plane waves
are incident along the horizontal direction when $p=0.1464$ m. The electric-field
distributions of (a) the line-transformed invisibility cloak with the PEC
inner boundary, (b) the PEC line. (c) The scattering width for the
line-transformed cloak, the PEC line, and the bare PEC cylinder.}

\vskip 5mm

\noindent \textbf{Fig. 6:} {(color online) TM-polarized plane waves
are incident horizontally when $p=0.1464$ m. The magnetic-field distributions of (a) the
line-transformed invisibility cloak with the PEC inner boundary, (b) the
PEC line. (c) The scattering width for the line-transformed cloak, the
PEC line, and the bare PEC cylinder.}

\newpage

\begin{figure}
\centerline{\includegraphics[width=11cm]{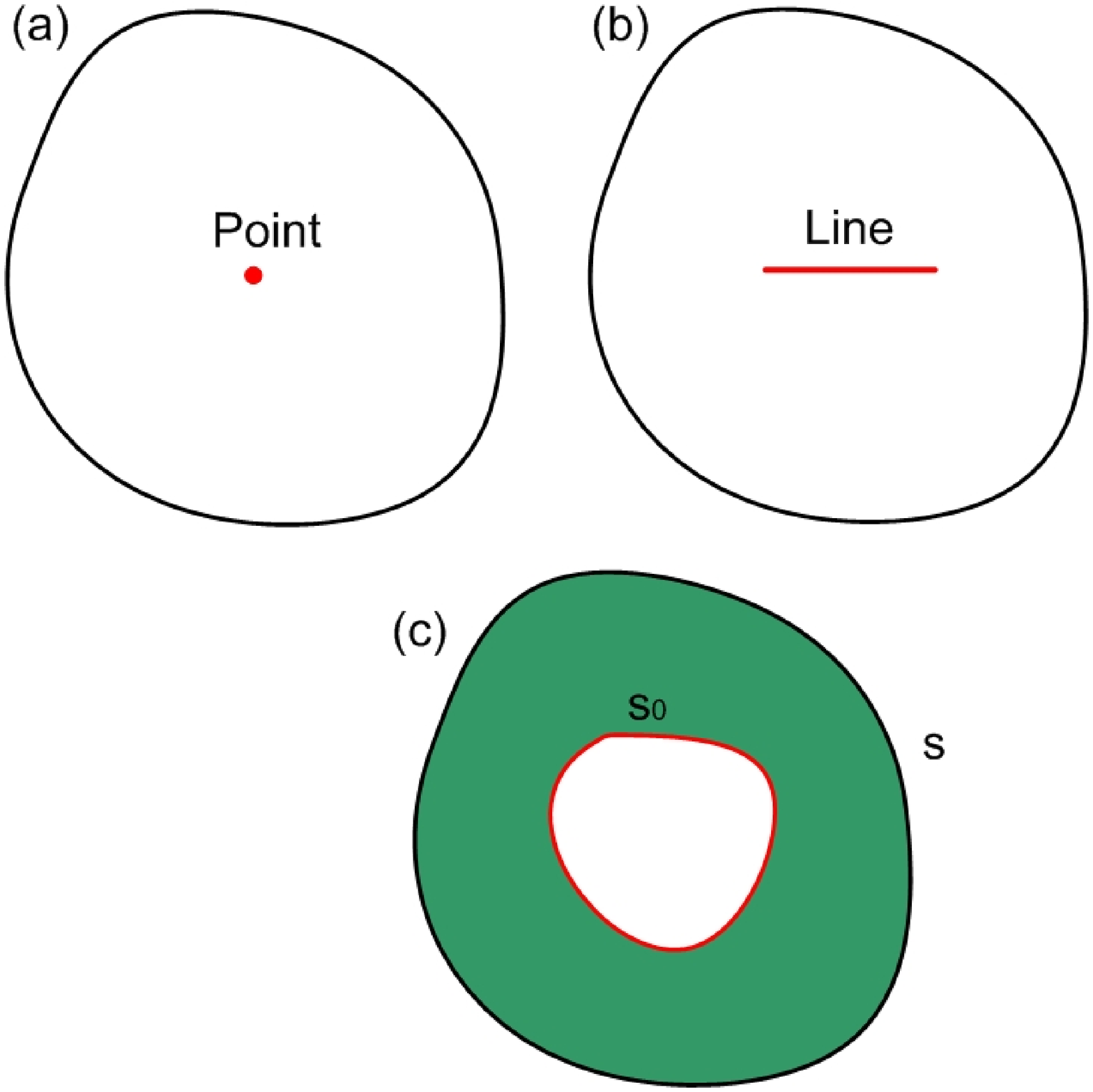}} \caption{}
\end{figure}

\begin{figure}
\centerline{\includegraphics[width=12cm]{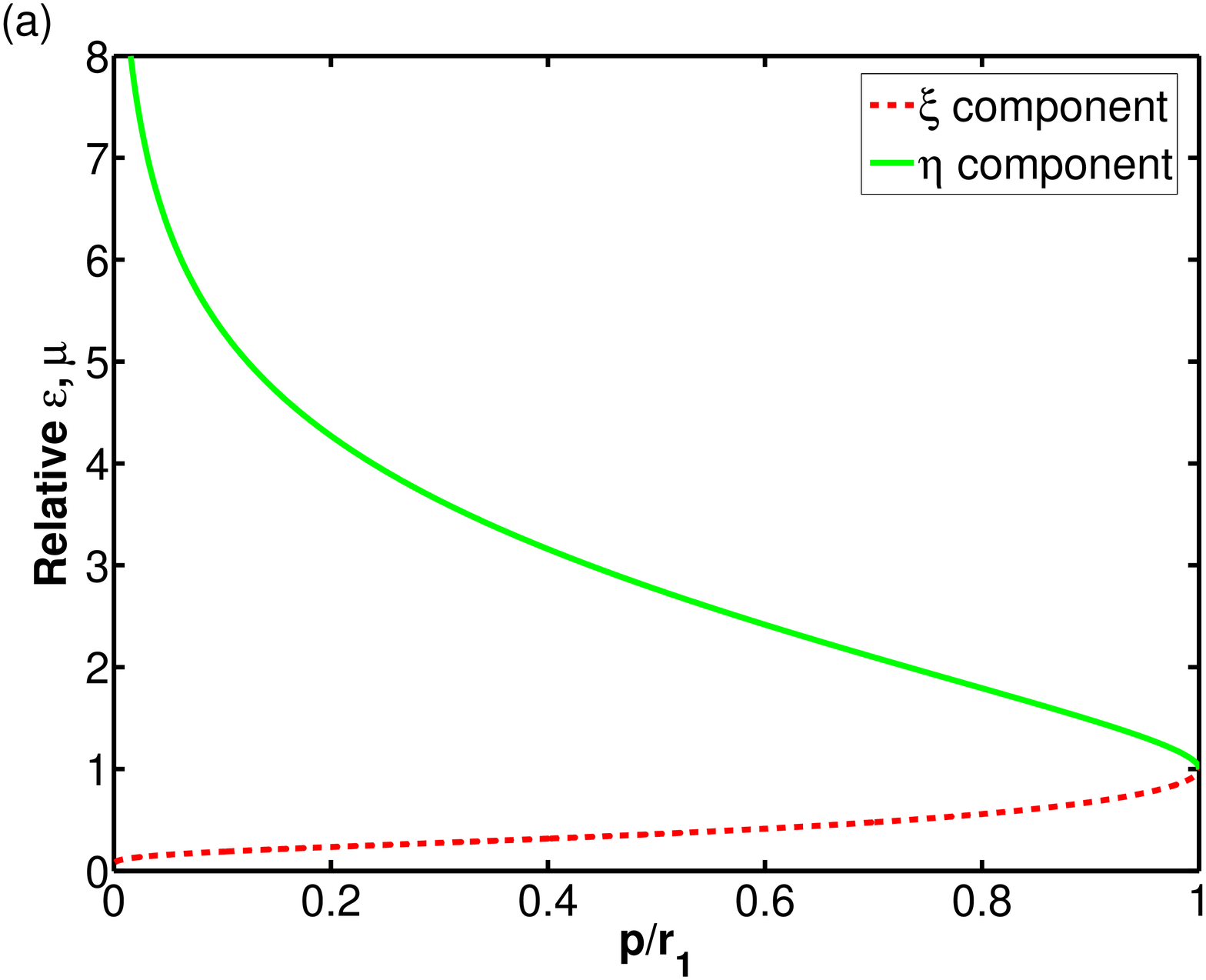}}
\centerline{\includegraphics[width=12cm]{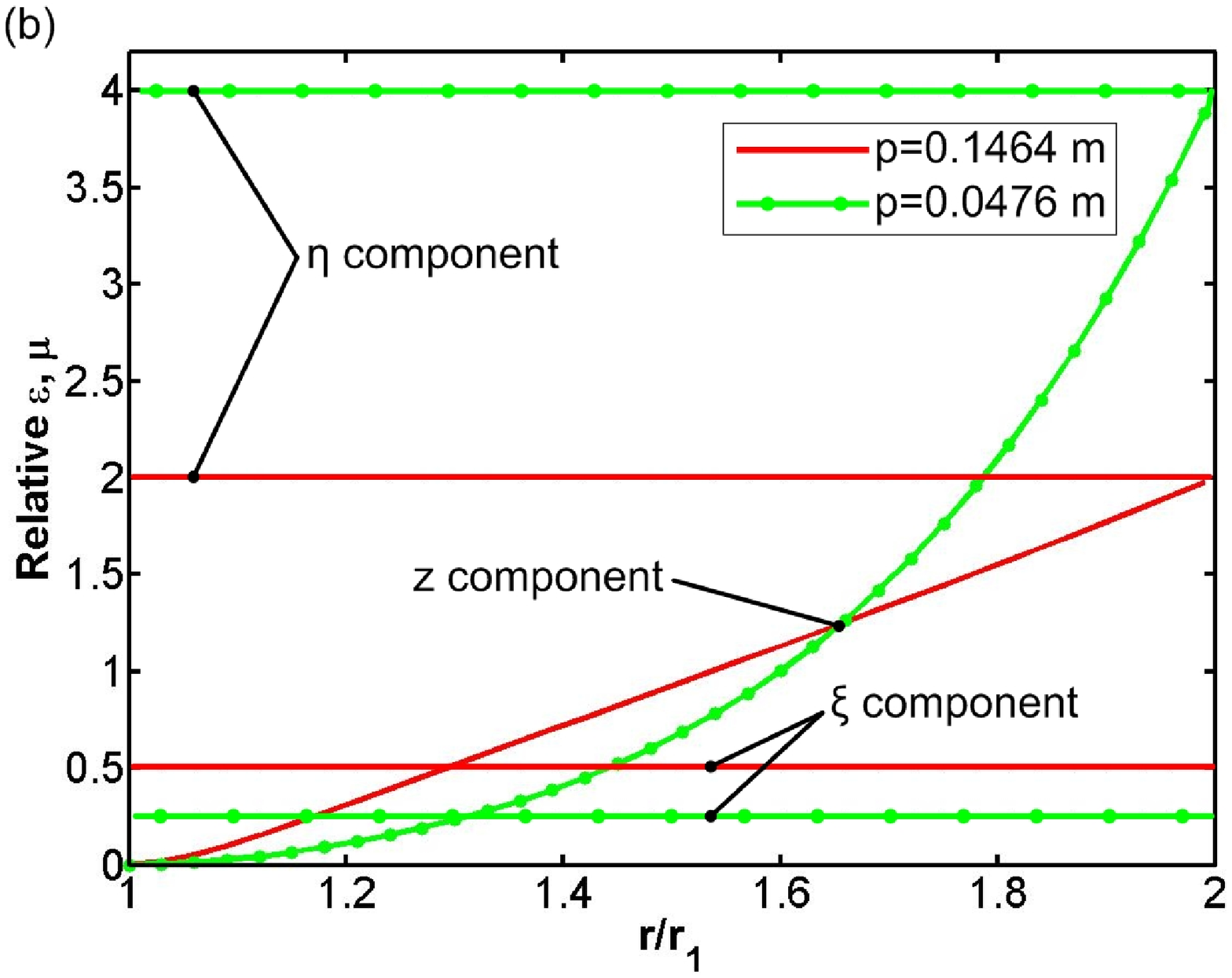}}
 \caption{}
\end{figure}

\begin{figure}
\centerline{\includegraphics[width=10cm]{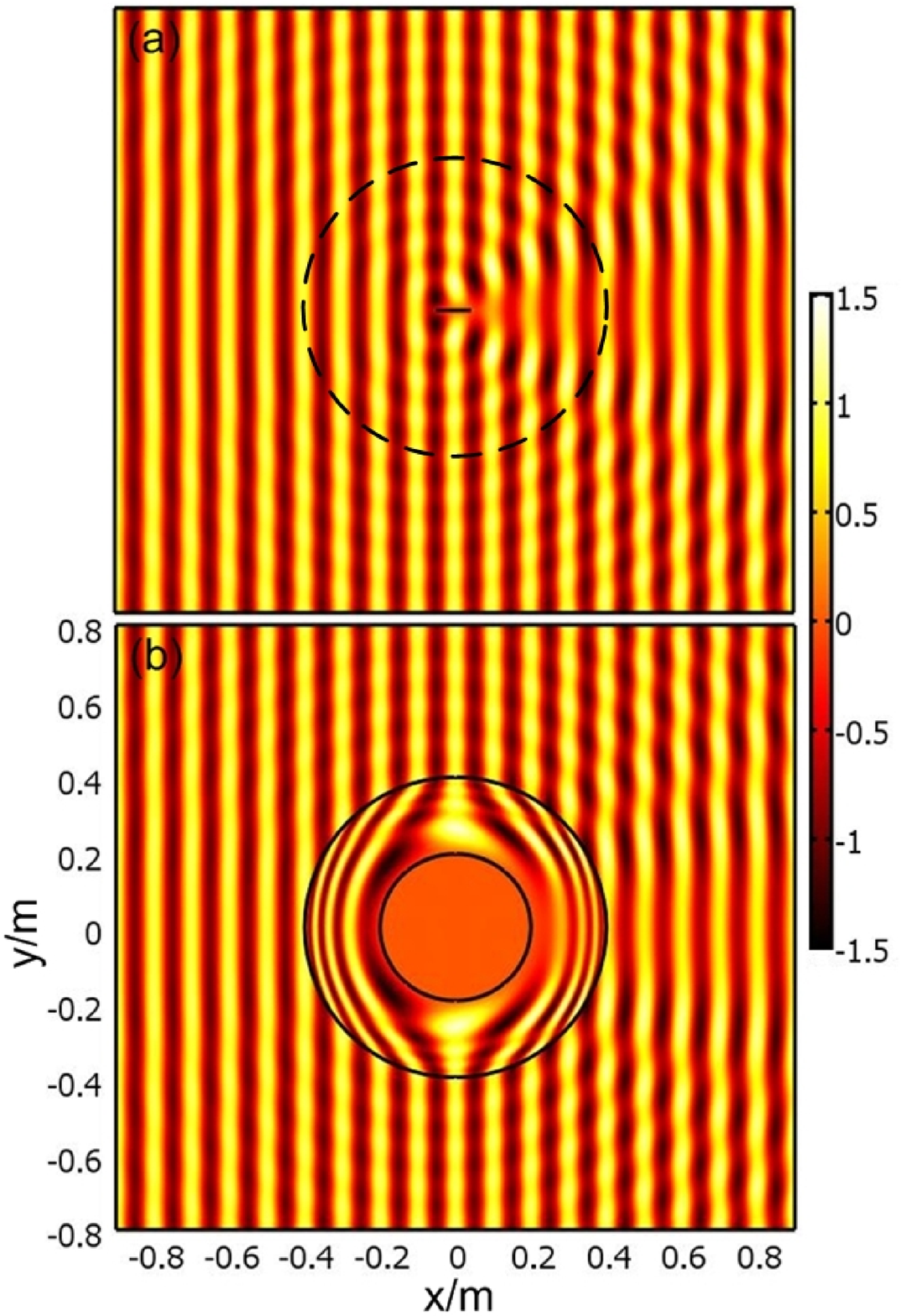}}
\centerline{\includegraphics[width=10cm]{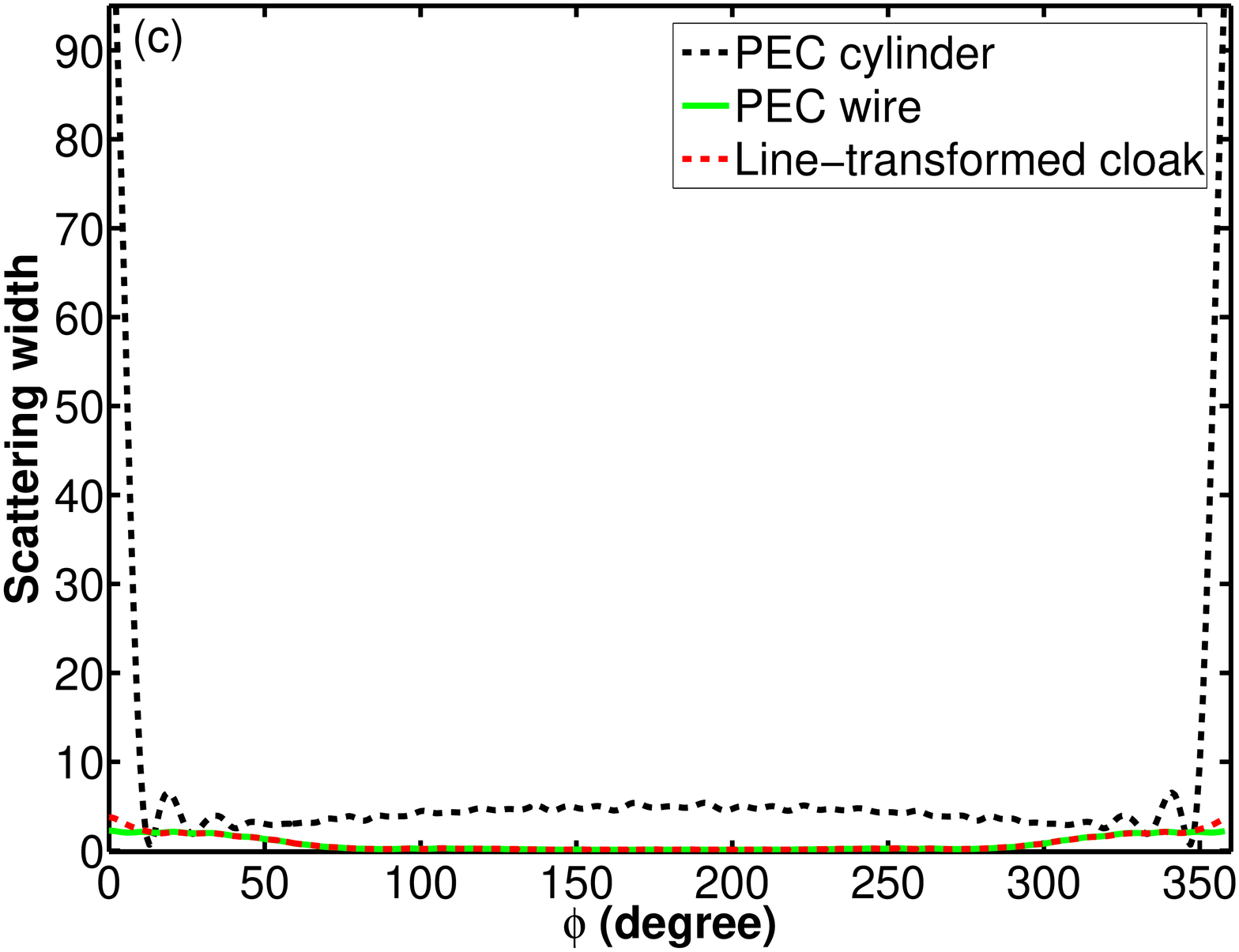}}
 \caption{}
\end{figure}

\begin{figure}
\centerline{\includegraphics[width=10cm]{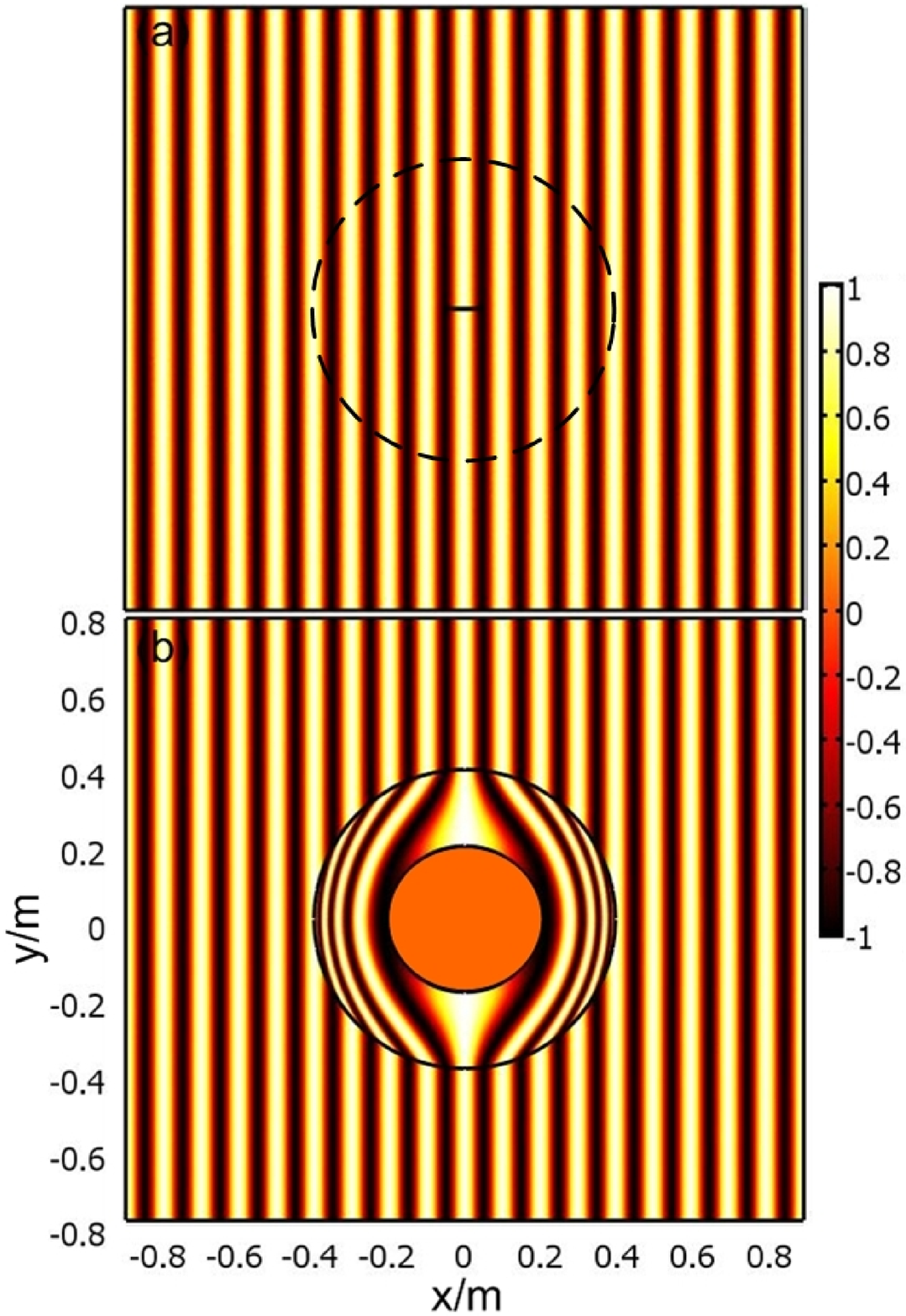}}
\centerline{\includegraphics[width=10cm]{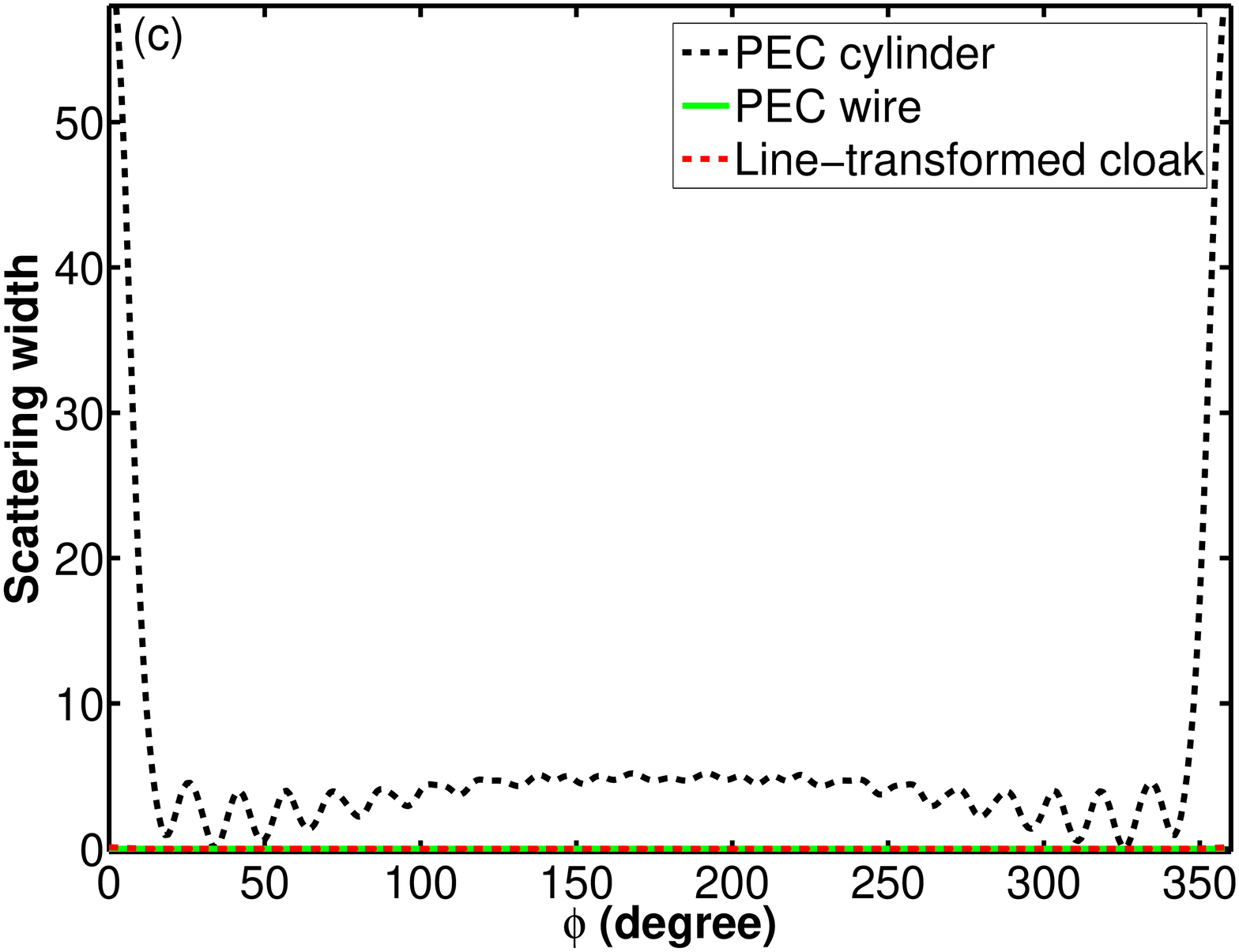}}
 \caption{}
\end{figure}

\begin{figure}
\centerline{\includegraphics[width=10cm]{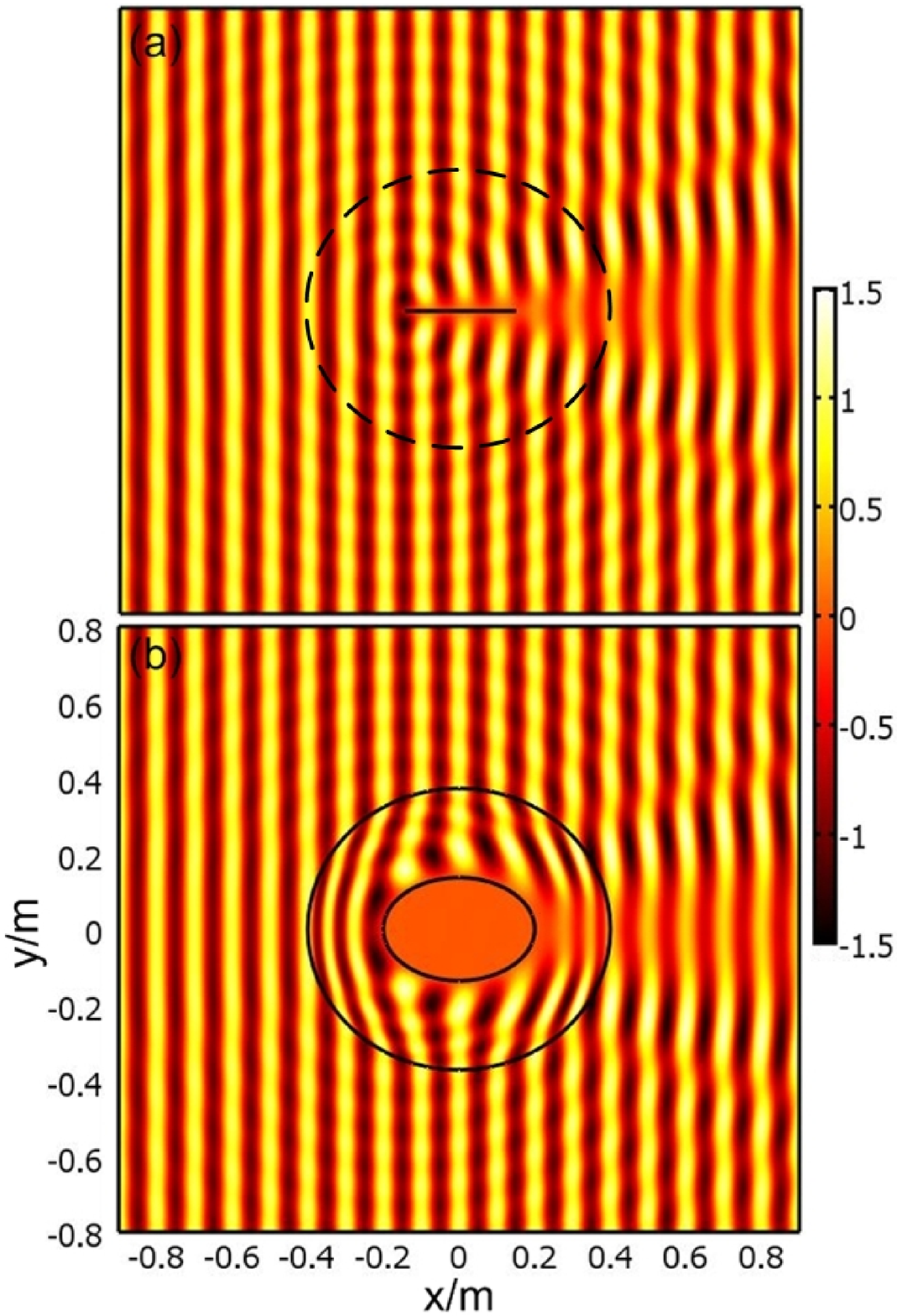}}
\centerline{\includegraphics[width=10cm]{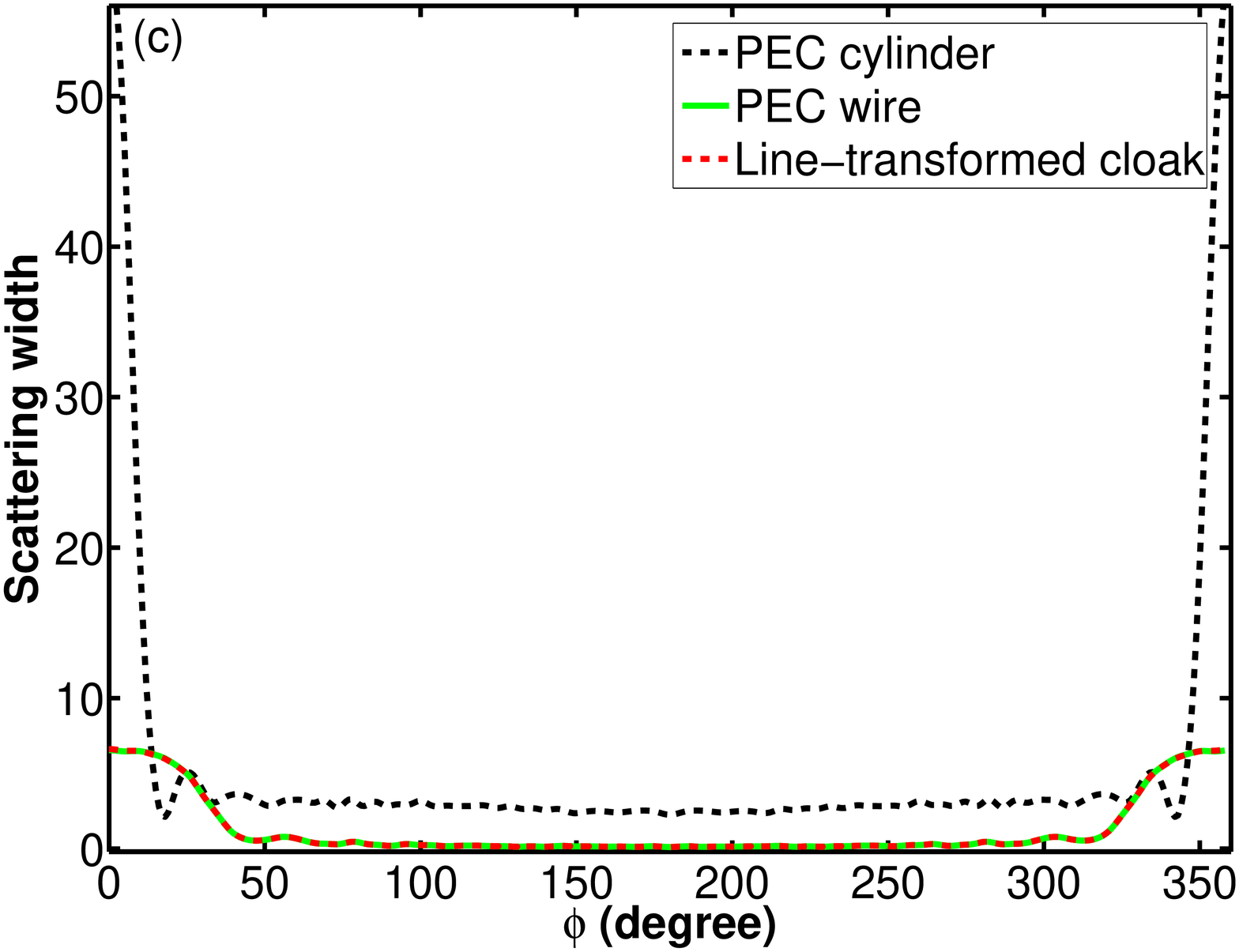}}
 \caption{}
\end{figure}

\begin{figure}
\centerline{\includegraphics[width=10cm]{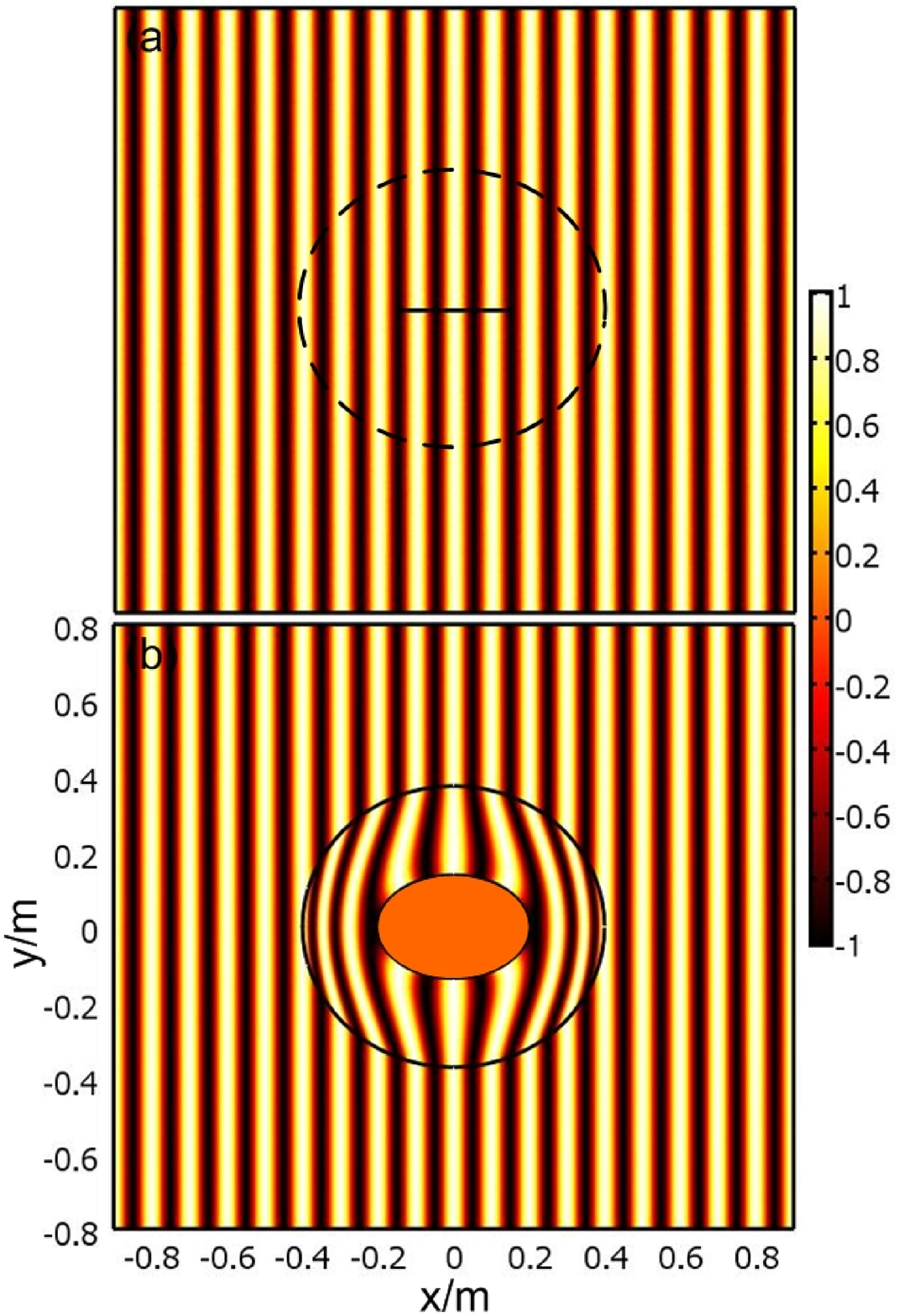}}
\centerline{\includegraphics[width=10cm]{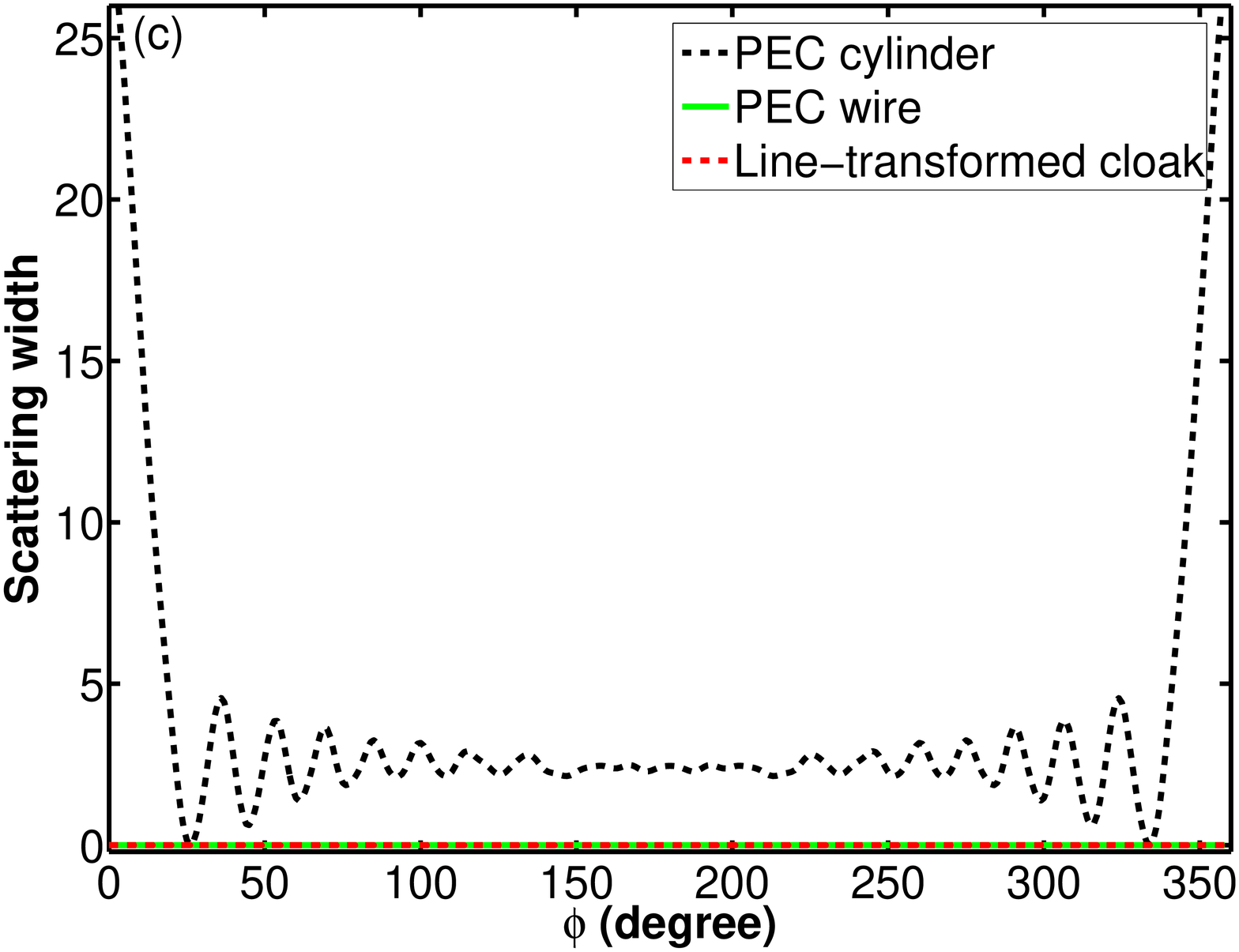}} \caption{}
\end{figure}

\end{document}